\documentclass[%
 reprint,
 amsmath,amssymb,
 aps,
]{revtex4-2}

\usepackage{graphicx}
\usepackage{dcolumn}
\usepackage{bm}

\usepackage{braket}
\usepackage[colorlinks,linkcolor=blue,anchorcolor=blue,citecolor=blue,urlcolor=blue]{hyperref}

\begin{document}

\preprint{APS/123-QED}

\title{Fragmented quantum phases in anti-blockade regime of Rydberg atom array}

\author{Han-Chao Chen$^{1,2}$}
\thanks{These authors contributed equally to this work.}
\author{Zheng-Yuan Zhang$^{1,2}$}
\thanks{These authors contributed equally to this work.}
\author{Meng Zhou$^{3}$}
\author{Xin Liu$^{1,2}$}
\author{Li-hua Zhang$^{1,2}$}
\author{Bang Liu$^{1,2}$}
\author{Lu-Xia Wang$^{3}$}
\author{Dong-Sheng Ding$^{1,2}$}
\email{dds@ustc.edu.cn}
\author{Bao-Sen Shi$^{1,2}$}

\affiliation{$^1$Laboratory of Quantum Information, University of Science and Technology of China, Hefei 230026, China.}
\affiliation{$^2$Anhui Province Key Laboratory of Quantum Network, University of Science and Technology of China, Hefei 230026, China.}
\affiliation{$^3$Department of Physics, Institute of Theoretical Physics, University of Science and Technology Beijing, Beijing 100083, China.}

\date{\today}

\begin{abstract}
In this Letter, we report a parameter-dependent Hilbert space fragmentation in a one-dimensional Rydberg atom array under anti-blockade conditions. We identify distinct non-equilibrium dynamical phases and show that their quasi-periodic behavior arises from the interplay of multi-path interference and multi-photon cascaded excitations, reflecting fundamental differences in state connectivity and effective dimensionality. By constructing a complete phase diagram, we reveal the transition from thermalization to fragmentation, and further demonstrate a secondary fragmentation process enabled by local constraints. Our results highlight the potential of anti-blockade mechanisms for programmable non-thermal dynamics in many-body systems.
\end{abstract}

\maketitle


\textit{Introduction.---} The eigenstate thermalization hypothesis (ETH) explains how isolated quantum systems thermalize without environmental coupling~\cite{1,2,3,4}. However, several mechanisms, including many-body localization~\cite{5,6,7}, quantum many-body scars~\cite{8,9,10}, and fracton phases~\cite{11,12,13}, have been shown to violate ETH and induce nonthermal behavior. Among these, Hilbert space fragmentation (HSF) has emerged as a distinctive mechanism that arises from intrinsic kinetic constraints, rather than disorder or conservation laws~\cite{14,15,16,17,18,19}. These constraints disconnect the Hilbert space into dynamically isolated subspaces, rendering some states inaccessible and preventing full thermalization. HSF was first identified in constrained models such as the PXP model~\cite{20,21}, constrained spin chains~\cite{22}, and dipole-conserving hopping models~\cite{23}, all exhibiting fragmented dynamics. Recent experiments using Rydberg atom arrays have directly observed such fragmentation~\cite{24,25,26}. Due to strong, controllable interactions and blockade effects~\cite{27,28,29,30}, Rydberg platforms naturally realize kinetic constraints, making them ideal for exploring HSF and nonthermal many-body dynamics. Beyond the standard blockade regime, anti-blockade conditions allow a fixed number of adjacent excitations while energetically forbidding others~\cite{31,32,33,34,35}. This introduces parameter-dependent local constraints that restructure the the connectivity of the accessible state space. While fragmentation may persist or take new forms, its stability, dynamical properties, and the boundaries between emergent sectors remain largely unexplored~\cite{36,37,38}.

\begin{figure*}[ht]
    \centering
    \includegraphics[width=1.0\linewidth]{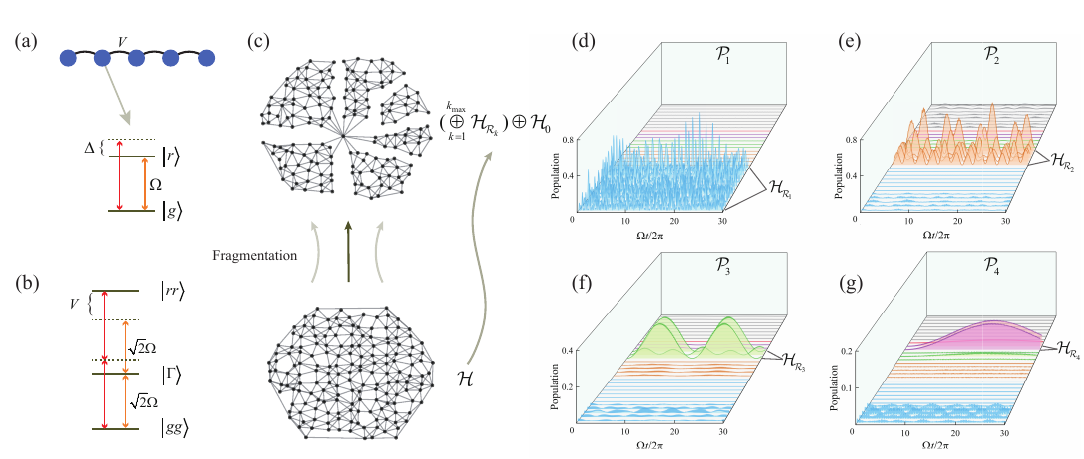}
    \caption{(a) Theoretical model. A one-dimensional array of five two-level atoms uniformly driven by a global laser field with Rabi frequency $\Omega$ and detuning $\Delta$. The red double-headed arrow marks $\Delta$, while the orange arrow denotes the resonant case. (b) Anti-blockade mechanism for two atoms. On resonance, interactions suppress double excitation (orange arrows). At $2\Delta=V$, a two-photon resonance enables $\ket{rr}$ excitation (red arrows). The state $\ket{\Gamma}=(\ket{rg}+\ket{gr})/\sqrt{2}$ represents a symmetric superposition. (c) Hilbert space fragmentation. Each dot represents a many-body state; solid lines link dynamically connected states. The full space $\mathcal{H}$ splits into disconnected subspaces ${\mathcal{H}_{\mathcal{R}_k}}$ sharing a common vacuum, plus a residual subspace $\mathcal{H}_0$ of inaccessible regions. (d)–(g) Dynamics within fragmented phases $\mathcal{P}_1$–$\mathcal{P}_4$. Time is in Rabi cycles (horizontal); state populations are on the vertical axis. Colored lines trace accessible states; gray lines mark inaccessible ones under anti-blockade constraints.}
    \label{fig:System}
\end{figure*}

In this work, we develop a theoretical model of a one-dimensional Rydberg atom array to investigate Hilbert space fragmentation induced by the anti-blockade effect. We first analyze the fragmentation of the system by simulating the dynamics of many-body excitations and identifying the structure of each subspace. Then, we systematically investigate the response of the system under different parameter regimes and construct a fragmented phase diagram to clarify the accessibility conditions and dynamical characteristics of each subspace. Finally, by introducing a local hard constraint, we investigate how the accessible subspace is further modified.

\begin{figure}[htpb]
    \centering
    \includegraphics[width=1.0\linewidth]{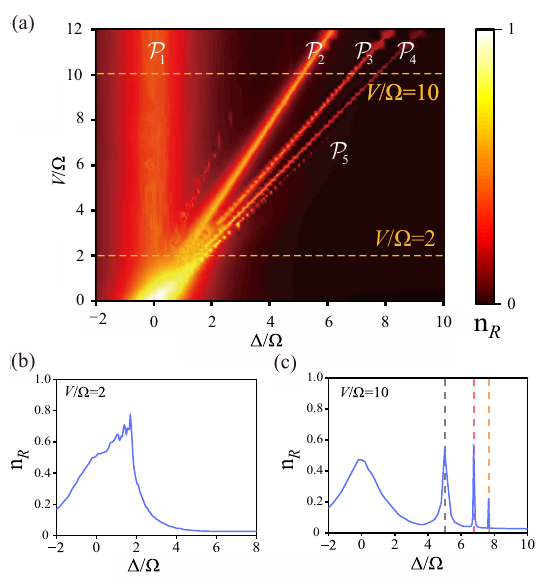}
    \caption{Fragmented quantum phase diagram. (a) The contour plot of the Rydberg excitation density $n_R$ in the $\Delta-V$ plane, showing the system’s response under different parameters. Labeled regions $\mathcal{P}_1$ through $\mathcal{P}_5$ denote fragmented quantum phases associated with distinct excitation subspaces. (b)(c) Rydberg excitation density as a function of detuning for $V/\Omega=3$ and $V/\Omega=10$. (b) In the weak interaction regime, the system satisfies the ETH, and all collective excited states are dynamically accessible. (c) Strong interactions lead to fragmented phases. The gray, red, and orange dashed lines indicate the resonance frequencies of $\mathcal{P}_2$, $\mathcal{P}_3$, and $\mathcal{P}_4$, respectively.}
    \label{fig:Spectrum}
\end{figure}

\textit{Theoretical model and Hilbert space fragmentation mechanism.---} Hilbert space fragmentation refers to the dynamical partitioning of the Hilbert space into disconnected subspaces due to local kinetic constraints which arise purely from the structure of allowed transitions in the many-body configuration space. In such systems, the unitary evolution remains confined to a restricted set of states, leading to non-ergodic behavior and preventing thermalization even in the absence of disorder. Our theoretical study focuses on a one-dimensional array consisted by equally spaced atoms, each modeled as a two-level system consisting of a ground state $\ket{g}$ and a Rydberg excited state $\ket{r}$. All atoms are subjected to a homogeneous light field, within the rotating wave approximation, the Hamiltonian governing the system can be expressed in the following form \cite{39,40,41,42,43,44}:
\begin{equation}
    \frac{\hat{H}}{\hbar}=\frac{\Omega}{2}\sum_i^N(\ket{g_i}\bra{r_i}+\ket{r_i}\bra{g_i})
    -\Delta\sum_i^Nn_i
    +V\sum_{i}^{N-1}n_in_{i+1},
\end{equation}
where $\Omega$ denotes the Rabi frequency, $\Delta$ represents the laser detuning, $V$ is the interaction strength between two nearest-neighbor Rydberg atoms, $N$ donates the number of atoms and the symbol $n_i=\ket{r_i}\bra{r_i}$ represents the Rydberg occupation operator at site $i$. The diagram of the system is shown in Fig.~\ref{fig:System}.(a). 

When the excitation light is near-resonant, interaction-induced energy shifts among Rydberg atoms give rise to the blockade effect, suppressing multiple excitations within a certain radius (orange double-headed arrows in Fig.~\ref{fig:System}(b)). In contrast, far-detuned multi-photon driving can compensate for these shifts, leading to anti-blockade effect, where neighboring atoms can resonantly excited via, e.g., two-photon processes (red arrows). This extends to $k$ adjacent excitations when the detuning matches the interaction energy. As a result, the system's accessible Hilbert space is constrained to configurations with at most $k$ consecutive excitations, enforcing local kinetic constraints and inducing Hilbert space fragmentation. To formalize this, we define $\mathcal{R}_0=0$ and introduce a set of kinetic projection operators $\mathcal{R}_k=S_k-\mathcal{R}_{k-1}$ that define the dynamically accessible subspaces for a given allowed run length $k$, where operator $S_k$ is defined as:
\begin{equation}
    S_k=1-\prod_{i=1}^{N}(n_i\prod_{j=1}^{k}n_{i+j})G(n_{i-1},n_{i+k}).
\end{equation}
In this expression, $G(n_{i-1},n_{i+k})=1-(1-n_{i-1})(1-n_{i+k})$. By defining $n_i=0$ when $i<1$ or $i>N$, $S_k$ acts as a projector that enforces the constraint that no more than $k$ adjacent atoms are simultaneously excited. Consequently, any state $\ket{\psi}$ in the Hilbert space satisfies:
\begin{equation}
    \mathcal{R}_k\ket{\psi} = 
\begin{cases}
\ket{\psi},& \text{when } \ket{\psi}\in \mathcal{H}_{\mathcal{R}_k}, 
\\ \ 0\ \ ,& \text{when } \ket{\psi}\notin \mathcal{H}_{\mathcal{R}_k}.
\end{cases}
\end{equation}
Here, $\mathcal{H}_{\mathcal{R}_k}$ denotes the dynamically accessible subspace corresponding to $\mathcal{R}_k$. As shown in Fig.~\ref{fig:System}(c), the full Hilbert space can be decomposed as: 
\begin{equation}
    \mathcal{H}=(\bigoplus_{k=1}^{k_{\text{max}}}\mathcal{H}_{\mathcal{R}_k})
    \bigoplus\mathcal{H}_0,
\end{equation}
where $\mathcal{H}_0$ consists of hybrid excited states that satisfy none of the $\mathcal{R}_k$ constraints (e.g., configurations such as $\ket{...rrgrg...}$ that violate all allowed excitation patterns).

To determine the parameters in our theoretical model, we base our choices on experimentally accessible values reported for the $^{87}$Rb systems. In particular, the controllable parameters can be tuned as follows: $\Omega=2$ MHz, $V=24$ MHz (Interaction strengths remain below this value in simulations), and neglect the next-nearest-neighbor interaction ($0.38$MHz)~\cite{8}, justifying the nearest-neighbor approximation. The detuning $\Delta$ could be widely tunable in experiments. Rydberg states with high principal quantum numbers can live over 100$\mu$s~\cite{45,46}, enabling $\sim$50 Rabi cycles; within the first 30, dissipation is negligible. We thus evolve observables using the quantum Liouville equation~\cite{47}, with all parameters scaled by $\Omega \equiv 1$,serving as a natural scale for comparison.

\begin{figure*}[htpb]
    \centering
    \includegraphics[width=1.0\linewidth]{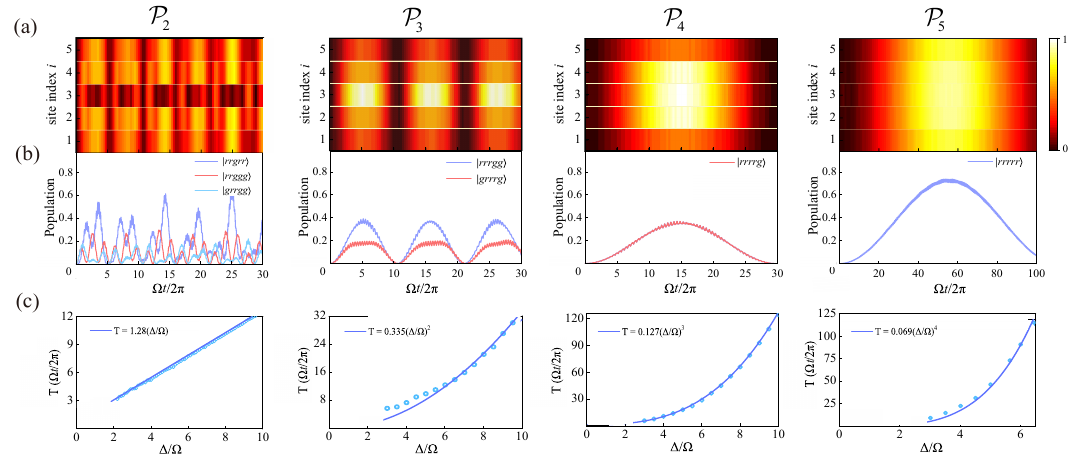}
    \caption{Dynamics within different fragmented quantum phases.
    (a) Heatmaps of Rydberg-state occupation dynamics for a five-atom system. The horizontal axis denotes time in units of Rabi cycles, and the vertical axis indicates site index. From left to right, the cases fr om $\mathcal{P}_2$ to $\mathcal{P}_5$ are shown in sequence. The color scale on the right represents the excitation density. (b) Time evolution of representative many-body basis states in each fragmented phases, showing quasi-periodic collective oscillations. (c) Dependence of the collective oscillation period $T$ on detuning $\Delta/\Omega$. Hollow blue circles represent numerically extracted periods under the anti-blockade resonance conditions for each fragmented phase. Solid blue curves are least-squares fits to the data, with the fitted functions labeled in each panel.}
    \label{fig:Dynamic}
\end{figure*}

\textit{Fragmented quantum phase diagram and dynamics.---} To systematically characterize the emergence of distinct excitation subspaces under anti-blockade conditions, we compute the phase diagram using a five-atom system, as shown in Fig.~\ref{fig:Spectrum}(a). Here, the color scale represents the peak of Rydberg excitation density $n_R=\sum_i^N\braket{n_i}/N$ over thirty Rabi oscillation cycles ($\Omega t/2\pi=30$). In the near-resonant regime, the $n_R$ exhibits a monotonic decrease with absolute value of detuning increases, ultimately reaching a regime where the array becomes nearly unresponsive to the excitation field. The two atom anti-blockade phenomenon emerges exclusively when the detuning and interaction satisfy $2\Delta\approx V$. As the detuning increases further, higher-order multi-photon excitations begin to emerge. Since only nearest-neighbor interactions are considered in the model, these higher-order excitations appear at positions satisfying $(N-1)V=N\Delta$. Unlike conventional quantum phases that vary continuously in parameter space, the Rydberg anti-blockade effect induced by interaction gives rise to a series of well-defined and isolated non-equilibrium dynamical phases $\{ \mathcal{P}_k \}$, which we refer to as fragmented quantum phases. However, when the interaction strength is relatively weak, the excitation branches corresponding to different fragmented quantum phase begin to overlap significantly, and the subspaces become connected. This phenomenon can be directly observed in the frequency response spectrum at $V/\Omega=2$ (Fig.~\ref{fig:Spectrum}(b)), where the weak interaction only induces slight energy level shifts and fails to impose strong local kinetic constraints, thereby preventing the system from exhibiting clear Hilbert space fragmentation.

As the interaction strength increases, the blockade and anti-blockade effects begin to compete, giving rise to a prominent crossover peak between the two regimes, where hybrid excited states become accessible. The eventual disappearance of this crossover feature with further increasing interaction strength marks a transition to a fully fragmented regime, in which the fragmented quantum phases become sharply separated in parameter space. The frequency response spectrum in the fully fragmented regime is shown in Fig.~\ref{fig:Spectrum}(c). The near-resonant blockade region and the far-detuned anti-blockade region are completely separated, with distinct resonance peaks appearing in the anti-blockade region corresponding to the $\mathcal{P}_2-\mathcal{P}_4$ phases. As the order increases, these peaks exhibit progressively narrower linewidths and reduced excitation probabilities. Such behavior is characteristic of multi-photon resonance processes, which are highly sensitive to both frequency and energy. Since the excitation must proceed through a series of intermediate virtual states, the energy matching conditions become increasingly stringent, resulting in enhanced frequency selectivity and significantly narrower effective linewidths with increasing photon number. Furthermore, since we fix the Rabi frequency to $\Omega\equiv 1$ in our calculations, the excitation probability, although formally scaling as $\Omega^{2k}$, does not increase with the order. Instead, it is suppressed by the cumulative detuning associated with the intermediate virtual states, leading to a rapid decrease in excitation efficiency~\cite{48,49}.

The heatmaps of the Rydberg excitation probability distributions across the five atomic sites for each phase $\mathcal{P}_2$ through $\mathcal{P}_5$ at a fixed interaction strength of $V/\Omega = 7$ are shown in Fig~\ref{fig:Dynamic}(a). It can be clearly observed that in each fragmented phase, the allowed excitation configurations strictly satisfy the corresponding projection operator $\mathcal{R}_k$, reflecting the underlying constraints that define each phase. Although these fragmented phases differ in their spatial excitation patterns, they share a set of common dynamical features: the system exhibits approximately adiabatic, non-diffusive evolution within each excitation sector and fails to explore the full Hilbert space, manifesting a key signature of Hilbert space fragmentation (Fig.~\ref{fig:System}(d)-(g)). Furthermore, we observe that the system exhibits prominent quasi-periodic revivals of collective excitation dynamics within the anti-blockade regime. To investigate the dynamical behavior of each fragmented quantum phase, we analyze the time evolution of several representative many-body states under resonance conditions (Fig.~\ref{fig:Dynamic}(b)) and extract the corresponding collective oscillation periods as functions of detuning $\Delta/\Omega$ (Fig.~\ref{fig:Dynamic}(c)). Our results show that the dependence of the oscillation period on detuning varies significantly across different fragmented phases. For example, in phase $\mathcal{P}_2$, the period increases approximately linearly with detuning, whereas in $\mathcal{P}_3$, a quadratic scaling behavior emerges. Interestingly, the polynomial power observed in the period–detuning relation tends to increase with the order $k$.

This process can be qualitatively understood in terms of multi-photon Raman transitions~\cite{50,51,52,53}. For a single quantum state, the effective multi-photon Rabi frequency follows the scaling $\Omega_{\text{eff}} \sim \Omega^k/\Delta^{k-1}$, where $\Delta$ is the detuning from the virtual intermediate state, i.e., the single-photon detuning. Since we fix $\Omega = 1$, the corresponding oscillation period scales as $T \sim (\Delta/\Omega)^{k-1}$. Within the fragmented subspace, the system dynamics are influenced by two primary effects. First, the existence of degenerate many-body excited states allows transitions to proceed through multiple quantum pathways. The differences in coupling strengths and relative phases among these pathways lead to quantum interference effects, which can enhance or suppress the overall excitation rate. Although such multipath interference introduces an additional factor that modifies the collective oscillation period, it does not alter the fundamental scaling relation between the detuning and the effective Rabi frequency. Second, non-degenerate many-body states can also be coupled through multi-photon processes, forming a cascaded excitation structure. In this case, beating effects between different energy levels give rise to quasi-periodic oscillations in the dynamics of system. While the collective quasi-period is determined by the least common multiple of the various beat frequencies, the dominant oscillation frequency is still governed by the effective Rabi frequency. This quasi-periodicity likewise does not disrupt the mathematical relation between the detuning and the oscillation period. These two effects coexist in the evolution of the system. Although they affect the observed oscillation period, neither breaks the fundamental scaling relationship between multi-photon transitions and the Rabi frequency. The systematic differences in scaling behavior clearly reflect underlying structural variations among the fragmented phases such as effective Hilbert subspace dimensionality, spectral connectivity, and the organization of accessible many-body states. These dynamical signatures serve as characteristic fingerprints for each fragmented quantum phase, offering an additional criterion for their identification and classification~\cite{54,55,56,57}.
\begin{figure}[htpb]
    \centering
    \includegraphics[width=1.0\linewidth]{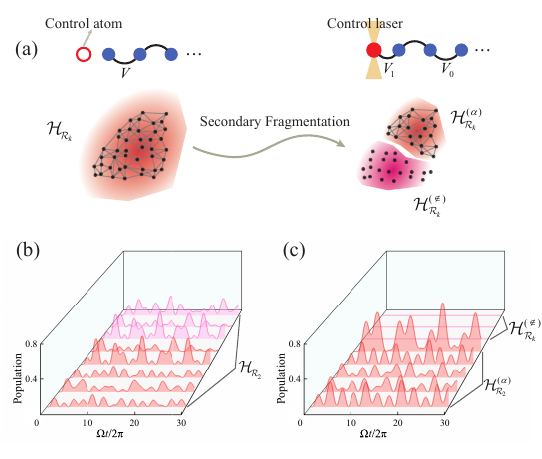}
    \caption{Secondary fragmentation. (a) Second fragmented system and its schematic: Red hollow/filled circles denote the control atom in ground/Rydberg state; blue filled circles represent atoms interacting via strength $V_0$. When excited, the control atom interacts more strongly ($V_1 > V_0$) with the first atom. The lower panel illustrates how the originally connected subspace becomes further fragmented due to the control atom. (b)(c) Dynamical evolution of states in subspace $\mathcal{H}_{\mathcal{R}_2}$ without (b) and with (c) secondary fragmentation control. Orange-red lines: states in $\mathcal{H}_{\mathcal{R}_k}^{(\alpha)}$; pink lines: states in $\mathcal{H}_{\mathcal{R}_2}^{(\notin)}$.}
    \label{fig:ControledExcitation}
\end{figure}

\textit{Secondary fragmentation within fragmented phase.---} Optical addressing techniques allow the precise excitation of individual atoms within an array~\cite{58,59}. Based on this capability, we can further reshape the set of dynamically accessible states within a fragmented subspace by exploiting the Rydberg blockade effect of the controllable atom. Specifically, when the control atom is in the ground state, the array can access all states that satisfy the $\mathcal{R}_k$ excitation conditions. However, once the control atom is excited to a Rydberg state, certain states in the Hilbert subspace $\mathcal{H}_{\mathcal{R}_k}$ become dynamically inaccessible. This leads to a further fragmentation of the subspace, which is now divided into two parts: $\mathcal{H}_{\mathcal{R}_k}^{(\alpha)}$, the accessible space, and $\mathcal{H}_{\mathcal{R}_k}^{(\notin)}$, the dynamically forbidden space. We refer to this phenomenon as secondary fragmentation. Unlike fundamental Hilbert space fragmentation, which partitions the accessible subspace based on symmetry or kinetic constraints, secondary fragmentation irreversibly reduces the accessible portion of the subspace. No tuning of system parameters can restore access to the confined region $\mathcal{H}_{\mathcal{R}_k}^{(\notin)}$. Figure \ref{fig:ControledExcitation}(a) illustrates this process,the control atom is positioned to the left of a six-atom chain and precisely adjusted such that only the first atom from the left lies within its blockade radius. Under the excitation conditions corresponding to the $\mathcal{P}_2$ phase, the dynamical evolution of all states in $\mathcal{H}_{\mathcal{R}_2}$ is shown in Fig.~\ref{fig:ControledExcitation}(b) when the control atom remains in the ground state. In contrast, when the control atom is excited to the Rydberg state via a $\pi$-pulse, states in $\mathcal{H}_{\mathcal{R}_2}^{(\notin)}$ are completely excluded from the dynamics, and those in $\mathcal{H}_{\mathcal{R}_2}^{(\alpha)}$ exhibit localized, non-diffusive evolution(Fig.~\ref{fig:ControledExcitation}(c)). Secondary fragmentation reveals how local constraints can reshape the global structure of state accessibility, imposing further limitations on the system's thermalization behavior.

\textit{Conclusion and outlook.---} This work systematically investigates many-body dynamics in a one-dimensional Rydberg atom array under the anti-blockade regime and reveals a parameter-dependent Hilbert space fragmentation induced by local excitation constraints. We show that under specific detuning, the anti-blockade effect permits limited-length contiguous excitations, restricting system evolution to specific subspaces. This fragmentation arises from dynamical constraints rather than Hamiltonian structure, and becomes increasingly pronounced with stronger interactions, leading to highly non-ergodic behavior. By scanning detuning and interaction strength, we identify multiple disconnected non-equilibrium dynamical phases and construct a complete quantum phase diagram, delineating the transition from thermalizing to fragmented regimes. Each phase exhibits quasi-periodic oscillations with distinct parameter dependencies, reflecting fundamentally different subspace connectivity and effective dimensionalities. Furthermore, we demonstrate secondary fragmentation by introducing local constraints via optical addressing, enabling finer control over accessible subspaces. Our results establish a versatile framework for programmable non-ergodic dynamics and open new avenues for exploring exotic non-equilibrium quantum phases~\cite{60,61}.

We acknowledge funding from the National Key R and D Program of China (Grant No. 2022YFA1404002), the National Natural Science Foundation of China (Grant Nos. T2495253, 61525504, 61435011).

\nocite{*}

\bibliography{citation}

\end{document}